\documentclass[twocolumn,preprint,3p,fleqn]{elsarticle}
\pdfoutput=1
\usepackage[hyperindex,breaklinks]{hyperref} 
\usepackage{atlasphysics}
\usepackage{amsmath}
\usepackage{subfigure}
\usepackage{comment}
\usepackage{graphicx}

\usepackage{preprintcover}
\PreprintCoverPaperTitle{Measurement of the mass difference between top and anti-top quarks in $pp$ collisions at $\boldmath{\sqrt{s}}$ = 7\,TeV using the ATLAS detector}
\PreprintIdNumber{CERN-PH-EP-2013-171} 
\PreprintCoverAbstract{A measurement of the mass difference between top and   
anti-top quarks is presented. In a 4.7\,fb$^{-1}$ data sample of proton--proton   
collisions at $\sqrt{s}$ = 7 TeV recorded with the ATLAS detector at   
the LHC, events consistent with $\ttbar$ production and decay into a   
single charged lepton final state are reconstructed. For each event, the   
mass difference between the top and anti-top quark candidate is   
calculated. A two $b$-tag requirement is used in order to reduce the background contribution. 
A maximum likelihood fit to these per-event mass   
differences yields $\dm \equiv \dmt = 0.67 \pm 0.61$(stat)$\pm 0.41 $(syst)\,GeV,
consistent with CPT invariance.}
\PreprintJournalName{Phys. Lett. B}

\def\dmt{\ensuremath{m_t-m_{\bar{t}}}}
\def\dm{\ensuremath{\Delta m}}
\def\dmf{\ensuremath{\Delta^{\mathrm{fit}}_m}}
\def\dmfi{\ensuremath{\Delta^{\mathrm{fit,i}}_m}}

\begin{document}
\begin{frontmatter}

\title{Measurement of the mass difference between top and anti-top quarks in $pp$ collisions at $\boldmath{\sqrt{s}}$ = 7\,TeV using the ATLAS detector}

\begin{abstract}
A measurement of the mass difference between top and   
anti-top quarks is presented. In a 4.7 fb$^{-1}$ data sample of proton--proton   
collisions at $\sqrt{s}$ = 7 TeV recorded with the ATLAS detector at   
the LHC, events consistent with $\ttbar$ production and decay into a   
single charged lepton final state are reconstructed. For each event, the   
mass difference between the top and anti-top quark candidate is   
calculated. A two $b$-tag requirement is used in order to reduce the background contribution. 
A maximum likelihood fit to these per-event mass   
differences yields $\dm \equiv \dmt = 0.67 \pm 0.61$(stat)$\pm 0.41 $(syst)\,GeV,
consistent with CPT invariance.
\end{abstract}
\end{frontmatter}

\section{Introduction}
The CPT symmetry~\footnote{CPT is the combination of three symmetries; Charge conjugation (C), Parity (P) and Time reversal (T).} required by a locally gauge-invariant quantum field theory dictates
that the masses of all particles and their anti-particles be exactly equal.
Any deviation from this would have major implications for
particle physics, implying a non-local field theory~\cite{cpt}.
Searches for CPT violation both in the $B$ meson sector~\cite{b1,b2,b3,b4} and with $K$ mesons~\cite{k1,k2,k3} have not 
yielded any deviations from the Standard Model (SM).
The top quark has the unique
property of decaying before hadronization, making it the only
quark for which a direct measurement of its mass is possible.
The CDF Collaboration measured the mass difference between 
top and anti-top 
quarks to be $\dm \equiv \dmt = 3.3 \pm 1.4 \pm 1.0$\,GeV~\cite{cdf}, approximately 
2 standard deviations away from zero. The D0 Collaboration measured $\dm= 0.8 \pm 1.8 \pm 0.5$\,GeV~\cite{d0}, in 
agreement with the SM value. The CMS Collaboration recently measured $\dm = -0.44 \pm 0.46 \pm 0.27$\,GeV~\cite{cms}, also in agreement with the 
SM value. 
The CDF and D0 analyses used both the top and anti-top quarks within each event to measure \dm. In the CMS measurement, the masses of the top and anti-top quarks with hadronic $W$ boson decays are extracted from two separate samples, split using the lepton charge, and subtracted from one another. In this Letter, the ATLAS Collaboration presents a measurement of this mass difference. The top and anti-top quarks are each taken from the same event, in which a $\ttbar$ pair is produced and decays in the lepton+jets channel.

\section{\bf{ATLAS detector}}
ATLAS~\cite{atlas} is a general-purpose particle physics detector with 
cylindrical geometry covering nearly the entire solid angle around the collision point.
Cylindrical coordinates $(r,\phi)$  are used in the transverse plane, where $\phi$ is the azimuthal
 angle around the beam pipe. The pseudorapidity is defined as $\eta\equiv-$ln tan$(\theta/2)$,
where $\theta$ is the polar angle. The 
transverse mass ($m_\mathrm{T}$) of any two objects is defined as $m_\mathrm{T} \equiv \sqrt{2 E_\mathrm{T}^1 E_\mathrm{T}^2 (1-\cos \Delta \phi)}$, where $E_\mathrm{T}$ is the object's transverse energy, defined in the plane transverse to the beam axis. 

The inner detector (ID) systems, located closest to the interaction region, are immersed in a 2~T axial magnetic field and provide
 charged particle tracking in the range $\vert\eta\vert<$~2.47. The ID systems 
consist of a 
high-granularity silicon pixel detector and a silicon microstrip tracker, as well as 
a transition radiation tracker. 
Located outside the solenoid, electromagnetic calorimetry is provided by barrel and endcap lead/liquid-argon 
calorimeters, and hadronic calorimetry by the steel/scintillating-tile sampling
calorimeters in the central region, and liquid-argon calorimeters in the endcap/forward regions. Comprising separate trigger and high-precision tracking chambers, the muon spectrometer measures
the deflection of muons in a magnetic field with a field integral from 2--8~T$\,{\mathrm{m}}$,
generated by one barrel and two endcap superconducting air-core toroids. 
A three-level trigger system is used to select and record interesting events. The level-1 hardware trigger uses a subset of detector information to reduce the event rate resulting from the peak LHC bunch crossing rate of 20 MHz in 2011 to a value of at most 65 kHz.
The level-1 trigger is followed by two software-based trigger levels, level-2 and the event filter, which together
reduce the event rate to a few hundred Hz for permanent storage and offline analysis.

\section{Data sample and event selection}
This analysis uses 4.7$\pm$0.2 fb$^{-1}$~\cite{b-lumi} of proton--proton collision data recorded by the ATLAS experiment at $\sqrt{s}$ = 7 TeV in 2011.
The selected events used in this analysis must contain the signature of a $\ttbar$ event decaying in the lepton+jets channel. 
Exactly one charged lepton is required -- either a single electron
with $\et > $ 25\,GeV, or a single muon with $\pt > $ 20\,GeV, where $\pt$ is the object's transverse momentum, defined in the plane transverse to the beam axis. 
Energy deposits are selected as electron candidates based on their shower shapes in the electromagnetic calorimeters and on the presence of a good-quality track pointing to them.
Electron candidates are required to pass the ``tight'' quality cuts described in Ref.~\cite{ref-electron}, 
to fall inside a well-instrumented region of the detector, and to be well isolated from other objects in the event.
Muons are required to pass ``tight''
muon quality cuts~\cite{b-mcp,b-mcp2,b-mcp3}, to be well measured in both the ID and the muon spectrometer, and to be isolated
from other objects in the event. 
Events with an electron (muon) are required to have been triggered by an electron (muon) 
trigger with an $\et$ ($\pt$) threshold of 20 (18)\,GeV. The selection requirements ensure that triggered events are on the trigger efficiency plateau~\cite{platE,platM}. 

Jets are reconstructed in the calorimeter using the anti-$k_t$ algorithm~\cite{b-akt,b-akt2}
with a radius parameter of 0.4, starting from energy deposits grouped into noise-suppressed topological clusters~\cite{Lampl:1099735,Barillari:1112035}. Jets are required to satisfy $\pt > $ 25\,GeV and $|\eta| <$ 2.5. Events
with jets arising from problematic regions in the calorimeters, beam backgrounds and cosmic rays are rejected~\cite{jetcorr}. Additional corrections are applied after the default ATLAS jet energy calibration~\cite{jetcorr} to restore on average the partonic energies in $\ttbar$ events.
Jets from the decay of long-lived heavy-flavor hadrons are selected by 
using a multivariate tagging algorithm ($b$-tagging) ~\cite{b-tagging,b-tagging2}.
The transverse momentum of neutrinos is inferred from the magnitude of the missing transverse momentum $(\met)$~\cite{b-MET}.

In addition to the requirement of exactly one charged lepton, the signal selection for this analysis requires four or more jets, at least two of which must be $b$-tagged.
 The selected lepton is required to match a trigger object that caused the event to be recorded. 
To suppress backgrounds from multi-jet events, $\met$ must be larger than 30 (20)\,GeV in the electron (muon) channel.
Further reduction of the multi-jet background in the electron channel is 
achieved by requiring the transverse mass ($m_\mathrm{T}$) of the lepton and $\met$ to be $>30$\,GeV. In the muon channel, $\met+m_\mathrm{T} >60$\,GeV is required.

\section{Simulated samples and background estimation}
The ATLAS detector simulation~\cite{simulation}, based on 
{\sc Geant4}~\cite{b-geant4}, is used to process simulated signal and background events. Simulated minimum bias collisions are overlaid on top of the hard scatter process, and events are reweighted so that the distribution of the average number of interactions (typically 5--20, see Ref.~\cite{pileup}) per bunch crossing matches the distribution observed in data. 

Simulated samples of $\ttbar$ events are produced using {\sc Pythia} v6.425~\cite{b-pythia} with $\dm$ ranging from $-15$\,GeV to $+15$\,GeV. In total, 15 such samples are used, with decreasing granularity at large $|\dm|$. Near $\dm=0$, the granularity is 0.3\,GeV~\footnote{In total, 15 signal samples were generated with a $\dm$ of $\pm15, \pm10, \pm5, \pm3, \pm1, \pm0.6, \pm0.3$ and $0$~GeV.}. In these samples, the average top quark mass ($\frac{m_t+m_{\bar{t}}}{2}$) is set to 172.5\,GeV. The underlying-event tune used is AUET2B~\cite{b-tune}, and the parton distribution function (PDF) set is MRST~\cite{b-mrst}. Despite being a leading order generator, {\sc Pythia} is used because it allows generation of events where the masses of the top and anti-top quarks are not equal. Non-zero widths as predicted by the SM for the corresponding top and anti-top quark masses are included in the event generation.

Pseudo-experiments and additional checks for systematic uncertainties are performed with a SM $\ttbar$ sample with $\dm=0$ generated using {\sc mc@nlo}~\cite{b-mcnlo,b-mcnlo2} v4.01 interfaced to {\sc Herwig} v6.520~\cite{b-herwig} and {\sc Jimmy} v4.31~\cite{b-jimmy}. Except for multi-jet processes, Monte Carlo simulations are used to study and estimate the backgrounds. The background from production of single $W$ bosons in association with jets is studied using {\sc Alpgen} v2.13~\cite{b-alpgen} interfaced to {\sc Herwig} and {\sc Jimmy}. The MLM matching scheme~\cite{b-mlm} is used to form inclusive $W$ + jets samples, taking appropriate care to remove overlapping events in heavy-flavor phase space stemming from both the hard scatter and the showering.  Diboson events are generated using {\sc Herwig}. Single-top events are generated using {\sc mc@nlo} in the $s-$ and $Wt-$channels, and {\sc AcerMC} v3.8~\cite{acer} in the $t$-channel. The distribution of the multi-jet background is taken from a control region in data where leptons are required to be semi-isolated and have large impact parameter ($d_0$) and impact parameter significance ($d_0$ divided by its uncertainty) with respect to the collision vertex. The semi-isolated selection requires the scalar $\pt$ sum for tracks in a cone of 0.3 around the electron (muon) divided by its $\et\,(\pt)$ to be between 0.1 and 0.3. The normalization of this background is obtained from a likelihood fit to the $\met$ distribution in data~\cite{QCD_fit}.

\section{Kinematic fits}
In order to measure a quantity sensitive to the mass difference $\dm$ between the top and anti-top quarks,
the kinematic $\chi^2$ fitter described below is used to reconstruct the $\ttbar$ system 
from the observed lepton, \met and jets. The assignment of the selected jets to the partons from the $\ttbar$ decay uses knowledge of the over-constrained $\ttbar$ system with the reconstructed top/anti-top quark mass difference ($\dmf$) as a free parameter in each event. 
In the kinematic fitter, the \pt of the lepton and jets
is allowed to fluctuate within uncertainties determined from simulated $\ttbar$ events. The average top quark mass is fixed, but the
individual $t$ and $\bar{t}$ masses are allowed to fluctuate while being constrained by the predicted top quark width. The masses of the two reconstructed $W$ bosons are also allowed to vary within the $W$ boson width. 
The fit is applied by examining
all jet--parton assignments (from among the five leading jets) consistent with the $b$-jet assignment and minimizing the following $\chi^2$:

\begin{eqnarray}
\label{}
\chi^2 & = & \sum_{i = \ell, 4 jets} \frac{\left(\pt^{i,\mathrm{fit}} - \pt^{i,\mathrm{meas}}\right)^2}{\sigma_i^2} \nonumber \\
       & + & \sum_{j = x,y} \frac{\left(p_j^{E_\mathrm{U}^{\mathrm{fit}}} - p_j^{E_\mathrm{U}^{\mathrm{meas}}}\right)^2}{\sigma_{E_\mathrm{U}}^2} \nonumber \\
       & + & \sum_{k=jj,\ell\nu}\frac{ \left(m_k^{\mathrm{fit}} - m_{W}\right)^2}{\sigma_{W}^2} \nonumber \\
       & + & \frac{\left(m_{b\ell\nu}^{\mathrm{meas}} - (m_{t} + \frac{(m_{b\ell\nu}^{\mathrm{fit}}  -  m_{bjj}^{\mathrm{fit}})}{2})\right)^2}{\sigma_t^2}\nonumber\\
       & + & \frac{\left(m_{bjj}^{\mathrm{meas}} - (m_{t} - \frac{(m_{b\ell\nu}^{\mathrm{fit}}  -  m_{bjj}^{\mathrm{fit}})}{2})\right)^2}{\sigma_t^2},
\end{eqnarray}
\noindent
\begin{eqnarray}
\label{deltaM}
\dmf & = & q_\ell \times (m_{b\ell\nu}^{\mathrm{fit}}  -  m_{bjj}^{\mathrm{fit}})
\end{eqnarray}
\noindent
where $\pt^{i,\mathrm{fit}}$ and $\pt^{i,\mathrm{meas}}$ are the fitted and measured $\pt$ of the jets and the charged lepton, and $\sigma_i$ is the uncertainty on those values.  The unclustered energy in the calorimeter ($E_\mathrm{U}$) is defined as a quantity that includes all energy
not associated with the 
primary lepton or the jets
 and is used to correct $\met.$  The width of the $W$ boson ($\sigma_W$) is set to the PDG value~\cite{pdg}, and the top quark width ($\sigma_t$) is set to the value predicted from theory.
The top quark mass ($m_t$) is fixed to 172.5\,GeV, and the $W$~boson mass ($m_W$) is set to $m_{W}=80.42$\,GeV. The value of $m_k^{\mathrm{fit}}$ is the fitted dijet (lepton--neutrino) mass from the hadronic (leptonic) $W$ boson decay, and 
$m_{b\ell\nu}^{\mathrm{fit}}$ and $m_{bjj}^{\mathrm{fit}}$ are the fitted top quark masses with leptonically and hadronically decaying $W$ bosons. 
The value of the mass difference between the hadronic- and leptonic-side top quarks is a free parameter in the fit. In each event, the single jet--parton assignment with the lowest $\chi^2$ is used,
and the fitted value of $\dmf$ is taken as an observable to measure the true $\dm$. As seen in Eq.~(\ref{deltaM}), $\dmf$ is calculated from the product of the lepton charge ($q_\ell$) and the difference between $m_{b\ell\nu}^{\mathrm{fit}}$ and $m_{bjj}^{\mathrm{fit}}$.  Events with $\chi^2 > 10$ for the best jet--parton assignment are considered to be poorly measured or background, and are rejected. The value of this cut is chosen based on studies of simulated signal events, and the efficiency of the $\chi^2$ selection is estimated in simulation to be 55\% for $\ttbar$ signal events and 31\% for background events. 
Table~\ref{tab:evts} shows the expected and observed number of events after all selection requirements, including the $\chi^2$ cut, are applied.

\begin{table}
\newcommand\T{\rule{0pt}{2.6ex}}
\newcommand\B{\rule[-1.2ex]{0pt}{0pt}}
\begin{center}
\begin{tabular}{ l  |  r@{\hspace{0.05cm}}l  | r@{\hspace{0.05cm}} l  }
\hline
Channel           & \multicolumn{2}{ c  |}{Muon} &  \multicolumn{2}{ c }{Electron} \\
\hline
Data              &  \multicolumn{2}{ c |}{$8854$}       & \multicolumn{2}{ c }{$4941$} \\ 
\hline 
\T SM $\ttbar \rightarrow W^+b W^-\bar{b}$   & $7700$ & $_{-1700}^{+1600}$ & $4500$ & $_{-1000}^{+900}$ \\
$W/Z+{\rm jets}$    & $320$ & $\pm90$ & $160$ & $\pm40$ \\
Single top  & $300$ & $\pm50$ & $170$ & $\pm30$ \\
Diboson  & $5$ & $\pm 1$ & $3$ & $\pm 1$ \\
Multi-jet      & $220$ & $\pm110$ & $110$ & $\pm60$   \\
\hline
\T \B Total expected (SM)& $8550$ & $_{-1700}^{+1600}$ & $4900$ & $_{-1000}^{+900}$ \\
\hline\hline
\end{tabular}
\caption{
The observed number of events in data, the expected numbers of events from signal and background processes and the total number of events, after all selection requirements.
Uncertainties shown include   
statistical and total systematic uncertainties added in quadrature.
\label{tab:evts}
}
\end{center}
\end{table}

Distributions of $\dmf$ are produced for all background samples as well as for a number of simulated
$\ttbar$ samples generated with different $\dm$. 

\begin{figure}[h!tb]
\includegraphics[width=\columnwidth]{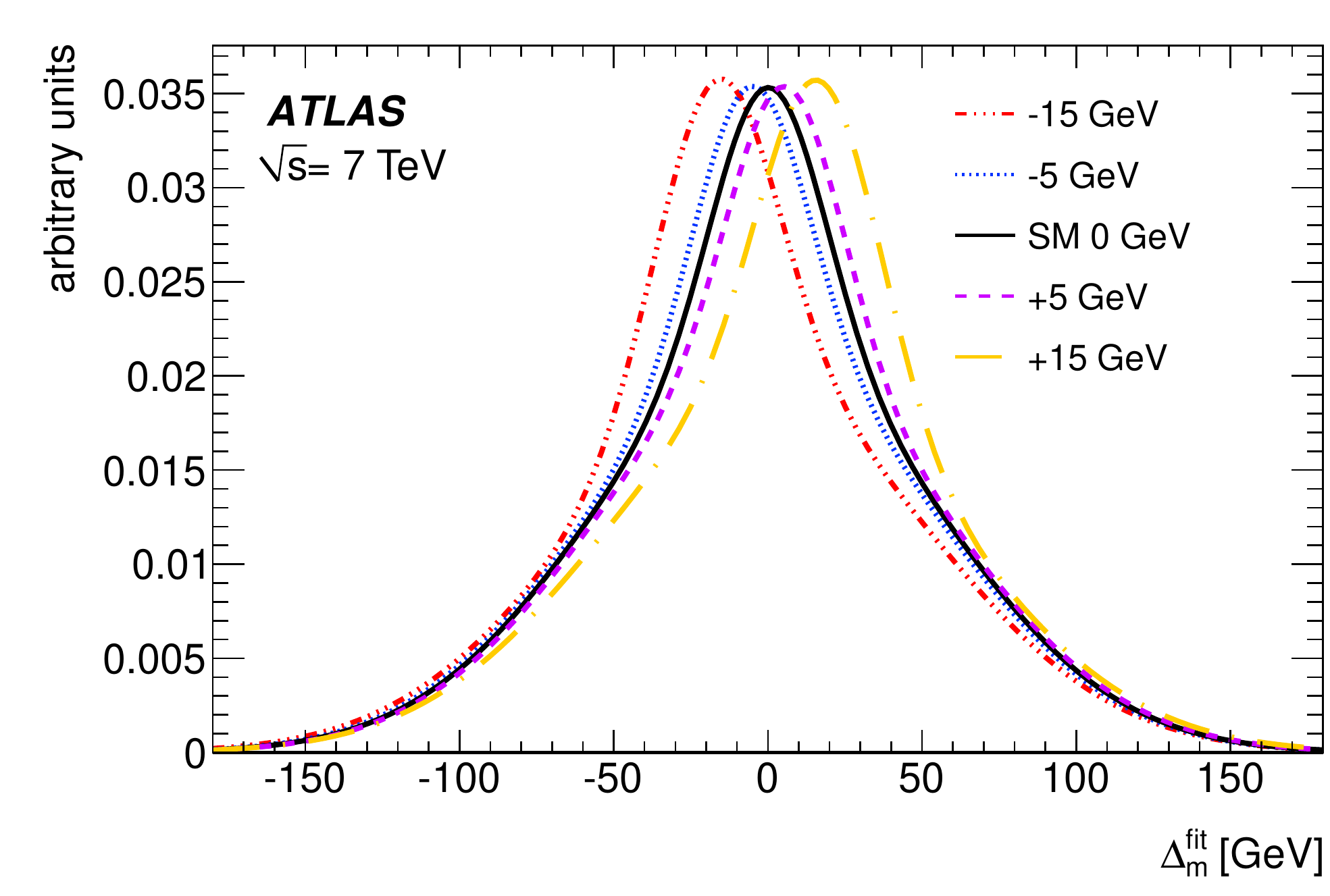}
   \caption{Parameterization of $\dmf$ for simulated $\ttbar$ samples   
with different values of \dm.
  \label{fig:paramsig}
  }
\end{figure}

\begin{figure}[h!tb]
  \includegraphics[width=\columnwidth]{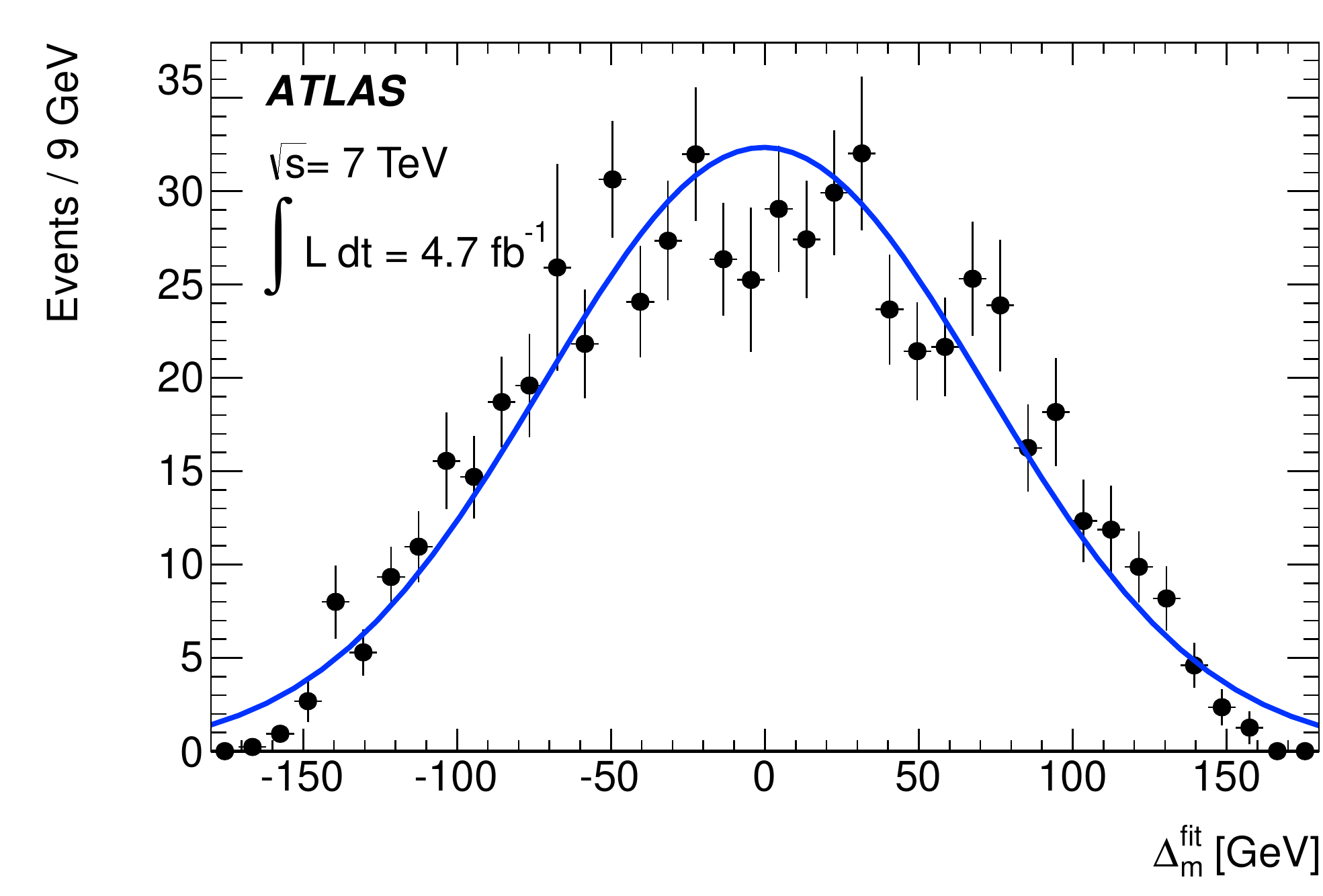}
   \caption[Background parameterizations]{Parameterized background template.
  \label{fig:background-param}
  }
\end{figure}

The $\dmf$ distributions in the signal samples are parameterized in templates  
by fitting the sum of two Gaussians, where the narrow one corresponds to
the correct jet--parton
pairing, and the wide one corresponds to an incorrect pairing. The widths of the two Gaussians are
quadratic functions of $\dm$ (symmetric about $\dm = 0$). The means of the 
two Gaussians are fit to linear functions of $\dm$. The relative weight of
the two Gaussians is fit to a quadratic function symmetric about $\dm = 0$.
Fig.~(\ref{fig:paramsig}) shows the parameterization for five different
values of $\dm$. The $\dmf$ distributions for all
background samples are combined with relative weights according to the SM prediction, into a single template distribution that is fit with a Gaussian, as shown in Fig.~(\ref{fig:background-param}). The choice of background parameterization has only a small impact on the
fits due to the small background in the double $b$-tag channel.
The signal and background templates are used to model the probability density distributions in $\dm$.

\section{Likelihood fit}

An unbinned extended maximum likelihood fit to the distribution of $\dmf$ is performed to extract $\dm$, as well as the 
expected number of signal ($n_s$) and background ($n_b$) events in the data. 
Given the data $D$, which contain $N$ values of $\dmf$, the probability distribution function for signal ($p_s$) and background ($p_b$) are used to write down a likelihood ($L$):

\begin{align}
\label{extendedmaximum}
L(D|&n_s, n_b, \dm) =\nonumber \\
&q(N, n_s + n_b) \times\nonumber \\
\prod_{i=1}^N &\frac{n_s p_s(\dmfi|\Delta_m) + n_b p_b(\dmfi)}{n_s+n_b}&
\end{align}
where $q(N, n_s + n_b)$ is the Poisson probability to observe $N$ events given $n_s+n_b$ expected events and the product over $i$ is over the $N$ reconstructed events. The likelihood is maximized over
all three parameters ($n_s$, $n_b$, $\dm$).
Ensembles of pseudo-experiments are run to ensure that the fits are unbiased and return correct statistical uncertainties.
The widths of pull distributions are consistent with unity. 
Due to the use of {\sc Pythia} to generate templates and {\sc mc@nlo} to run ensemble tests, a 175~\MeV offset is applied to all pseudo-experiments (and to the nominal fit result) to return an unbiased measurement, with the statistical uncertainty of 50~\MeV on this calibration taken as a systematic uncertainty. The 175~\MeV offset is the average difference between the {\sc mc@nlo} samples with the top and anti-top quark masses reweighted to the distributions in {\sc pythia} for a given mass difference. 
When running pseudo-experiments, events are drawn directly from the simulated samples and not from the 
parameterizations in order to check for any potential bias.

\begin{figure}
\begin{center}
\includegraphics[width=\columnwidth]{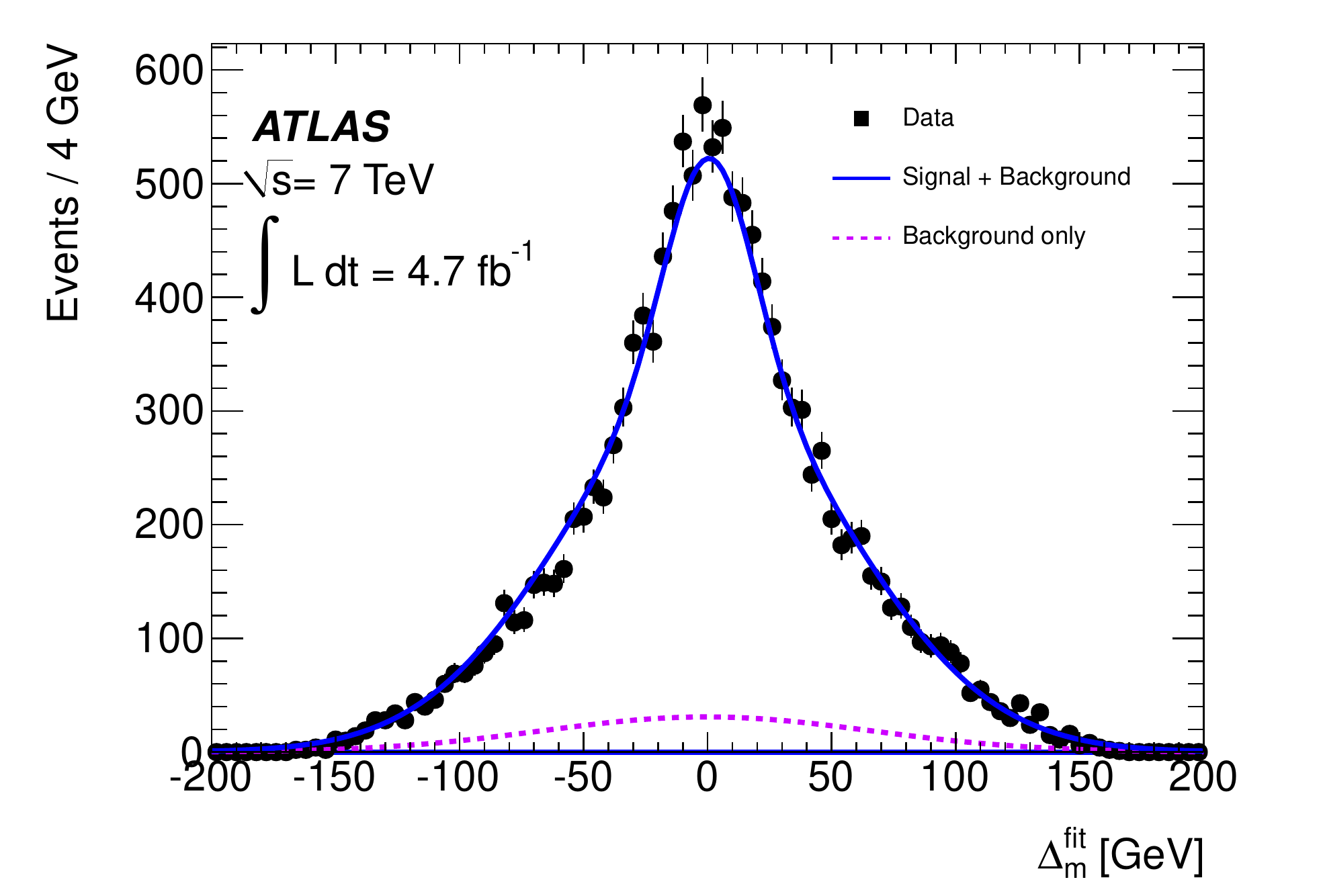}
\caption{ \label{fig:MDdata}
Reconstructed top/anti-top quark mass difference with the best maximum likelihood fit for signal and background overlaid.}
\end{center}
\end{figure}

The extended maximum likelihood fit is applied to the full 2011 dataset, yielding the result shown in Fig.~(\ref{fig:MDdata}). The value of 175 MeV quoted above is subtracted from the result to correct for this bias, giving a measured top/anti-top quark mass difference of $\dmt = 0.67 \pm 0.61$(stat). The $\chi^2$ per degree of freedom for Fig.~(\ref{fig:MDdata}) is 1.2. 

\begin{table}[h!]
\begin{center}
\begin{tabular}{|l|c|c|}
\hline
Systematic Uncertainty               &  $\Delta (\dm)$ [GeV] \\
\hline\hline
$b/\bar{b}$ decay uncertainties & 0.34 \\
$K^+/K^-$ calorimeter & \\
\,\,\,\,\,\,\,\, response asymmetry & 0.08 \\
Residual $b$ vs $\bar{b}$ differences & 0.08 \\
\hline
$b$-tagging                	       & $ 0.08$ \\ 
Mis-tagging as a $b$-quark jet     	& $ 0.05$ \\
\hline
Jet energy scale                       &  $ 0.04$  \\ 
$b$-jet energy scale                     &  $ 0.05$ \\
Jet energy resolution                  &  $ 0.03$  \\
\hline
Parton shower                          & $ 0.08$ \\
MC generator                           &  $ 0.08$ \\
ISR/FSR                                &  $ 0.07$ \\
\hline
Calibration method               & $ 0.05$ \\ 
Non-$\ttbar$ normalization               & $ 0.04$ \\ 
Non-$\ttbar$ shape               & $ 0.04$ \\ 
\hline
Parton distribution function               & $ 0.02$ \\ 
Lepton energy scale asymmetry                 & $ < 0.01 $ \\
Electron reconstruction&\\
\,\,\,\,\,\,\,\, \& identification          &  $ 0.02$ \\ 
Muon reconstruction&\\
\,\,\,\,\,\,\,\, \& identification               & $ 0.04$ \\ 
Top mass input              & $ 0.04$ \\ 
\hline\hline
Total               & $ 0.41$ \\ 
\hline
\end{tabular}
\caption{\label{tab:systeffect}
Systematic uncertainties.}
\end{center}
\end{table}

\section{Systematic uncertainties}
Due to cancellations from measuring the mass difference and not the individual quark masses, most systematic effects yield small uncertainties on the final measurement.
Systematic uncertainties are evaluated by performing pseudo-experiments with pseudo-data that reflect a variation due to the potential source of uncertainty considered, and comparing the extracted $\dm$ to the one obtained with default pseudo-data.
A list of all systematic uncertainties and their effects on the measurement are summarized in Table~\ref{tab:systeffect}. The total systematic uncertainty of 0.41 GeV on the measured $\dm$ is dominated by the uncertainty from the choice of $b$ fragmentation model, which can induce different detector response to jets from $b$- and $\bbar$-quarks in simulation. The various systematic uncertainties are discussed in more detail below.

Systematic uncertainties on $\dm$ due to differences in the detector response to jets from $b$- and $\bbar$-quarks are difficult to evaluate with in-situ methods in the $\ttbar$ environment due to correlations with $\dm$.
Based on the evaluation of the jet energy scale uncertainty from single-hadron response measurements~\cite{singleresponse}, most differences between the calorimeter response to the two types of jets are expected to be small; exceptions are discussed below. One such difference could come from the different responses to positively and negatively charged kaons, which occur at different rates in jets from $b$- and $\bbar$- quarks. The interaction cross sections for $K^+$ and $K^-$ in the calorimeters are different. Such effects are studied by comparing convolutions of the kaon spectra in $b$- and $\bbar$-jets from $\ttbar$ events with the expected calorimeter response to kaons simulated with various hadron shower simulation models, as specified in Ref.~\cite{singleresponse}. The resulting uncertainty is 80 MeV.
Uncertainties due to fragmentation and the decay of $b$-hadrons can also  
lead to uncertainties in the particle content and hadron   
momentum spectra, and thus in the calorimeter response. This uncertainty 
is evaluated by comparing {\sc powheg} samples that use {\sc Evtgen}~\cite{evtgen} and {\sc Pythia} to decay $b$-hadrons, and is estimated to be 340 MeV. The {\sc Evtgen} particle decay simulation implements different hadron decay models and up-to-date $b$-hadron decay tables. An additional 80 MeV is assigned to account for
any residual difference in response between jets from $b$ and $\bbar$ quarks due to effects not considered above. Parton shower and additional fragmentation uncertainties are estimated by comparing {\sc Powheg} samples interfaced with {\sc Herwig} to those interfaced with {\sc Pythia}. 

Other uncertainties are small compared to those from differences between jets from $b$- and $\bar{b}$-quarks. The uncertainty on $\dm$ from the  uncertainty on the $b$-tagging efficiency is measured by varying the $b$-tag scale factors, which correct simulated efficiencies to those measured in data, within 1$\sigma$ of their uncertainties. 
The systematic effects from uncertain light- and $b$-jet energy scales and resolutions are small, as they 
affect the top and anti-top quark masses in the same way~\cite{JES,bJES}.
Generator uncertainties are estimated by comparing pseudo-experiments using {\sc mc@nlo} and {\sc Powheg}. 
A systematic uncertainty on the amount of QCD radiation is derived from {\sc AcerMC} $\ttbar$ samples 
that have varying amounts of initial- and final-state radiation~\cite{radiation}.
Uncertainties from the template parameterization are estimated by varying the parameters
within their uncertainties, and are found to be small.
The systematic uncertainties due to background shape and rate
 are estimated by replacing the $W$+jets background used in pseudo-experiments with the shape from the multi-jet background and by varying the normalization within uncertainties. A small systematic uncertainty due to the 
parton distribution functions of the proton is evaluated by taking the envelope of the MSTW2008NLO~\cite{mstw}, NNPDF2.3~\cite{nnpdf} and 
CTEQ6.6~\cite{ct66} PDF set uncertainties, following the 
PDF4LHC recommendations~\cite{pdf4lhc}.
Asymmetries due to lepton energy scales are negligible. A systematic uncertainty on the top quark mass of 40\,MeV is estimated by comparing pseudo-experiments where the input average top quark mass is shifted up and down by 1.5\,GeV. Other systematic uncertainties considered are those caused by the uncertainty on the lepton identification and reconstruction.

\section{Conclusions}
The analysis described in this Letter is the first measurement by ATLAS of the mass difference between the top and anti-top quarks using event-by-event quantities in $\ttbar$ events. It is based on 4.7~fb$^{-1}$ of 7~TeV proton--proton collisions at the LHC. The mass difference, $\dm$, is calculated using a kinematic $\chi^2$ fitter. 
The measured mass difference is $\dm \equiv \dmt = 0.67 \pm 0.61$(stat) $\pm 0.41 $(syst)\,GeV, consistent with the SM expectation of no mass difference. 

\section{Acknowledgements}

We thank CERN for the very successful operation of the LHC, as well as the
support staff from our institutions without whom ATLAS could not be
operated efficiently.

We acknowledge the support of ANPCyT, Argentina; YerPhI, Armenia; ARC,
Australia; BMWF and FWF, Austria; ANAS, Azerbaijan; SSTC, Belarus; CNPq and FAPESP,
Brazil; NSERC, NRC and CFI, Canada; CERN; CONICYT, Chile; CAS, MOST and NSFC,
China; COLCIENCIAS, Colombia; MSMT CR, MPO CR and VSC CR, Czech Republic;
DNRF, DNSRC and Lundbeck Foundation, Denmark; EPLANET, ERC and NSRF, European Union;
IN2P3-CNRS, CEA-DSM/IRFU, France; GNSF, Georgia; BMBF, DFG, HGF, MPG and AvH
Foundation, Germany; GSRT and NSRF, Greece; ISF, MINERVA, GIF, DIP and Benoziyo Center,
Israel; INFN, Italy; MEXT and JSPS, Japan; CNRST, Morocco; FOM and NWO,
Netherlands; BRF and RCN, Norway; MNiSW, Poland; GRICES and FCT, Portugal; MNE/IFA, Romania; MES of Russia and ROSATOM, Russian Federation; JINR; MSTD,
Serbia; MSSR, Slovakia; ARRS and MIZ\v{S}, Slovenia; DST/NRF, South Africa;
MINECO, Spain; SRC and Wallenberg Foundation, Sweden; SER, SNSF and Cantons of
Bern and Geneva, Switzerland; NSC, Taiwan; TAEK, Turkey; STFC, the Royal
Society and Leverhulme Trust, United Kingdom; DOE and NSF, United States of
America.

The crucial computing support from all WLCG partners is acknowledged
gratefully, in particular from CERN and the ATLAS Tier-1 facilities at
TRIUMF (Canada), NDGF (Denmark, Norway, Sweden), CC-IN2P3 (France),
KIT/GridKA (Germany), INFN-CNAF (Italy), NL-T1 (Netherlands), PIC (Spain),
ASGC (Taiwan), RAL (UK) and BNL (USA) and in the Tier-2 facilities
worldwide.

\bibliographystyle{atlasBibStyleWoTitle}
\bibliography{paper_v1}

\onecolumn
\clearpage 
\begin{flushleft}
{\Large The ATLAS Collaboration}

\bigskip

G.~Aad$^{\rm 48}$,
T.~Abajyan$^{\rm 21}$,
B.~Abbott$^{\rm 112}$,
J.~Abdallah$^{\rm 12}$,
S.~Abdel~Khalek$^{\rm 116}$,
O.~Abdinov$^{\rm 11}$,
R.~Aben$^{\rm 106}$,
B.~Abi$^{\rm 113}$,
M.~Abolins$^{\rm 89}$,
O.S.~AbouZeid$^{\rm 159}$
% ...
\bigskip
\\
% ...
$^{11}$ Institute of Physics, Azerbaijan Academy of Sciences, Baku, Azerbaijan\\
$^{12}$ Institut de F{\'\i}sica d'Altes Energies and Departament de F{\'\i}sica de la Universitat Aut{\`o}noma de Barcelona, Barcelona, Spain\\
% ...
$^{21}$ Physikalisches Institut, University of Bonn, Bonn, Germany\\
% ...
$^{48}$ Fakult{\"a}t f{\"u}r Mathematik und Physik, Albert-Ludwigs-Universit{\"a}t, Freiburg, Germany\\
% ...
$^{89}$ Department of Physics and Astronomy, Michigan State University, East Lansing MI, United States of America\\
% ...
$^{106}$ Nikhef National Institute for Subatomic Physics and University of Amsterdam, Amsterdam, Netherlands\\
% ...
$^{112}$ Homer L. Dodge Department of Physics and Astronomy, University of Oklahoma, Norman OK, United States of America\\
$^{113}$ Department of Physics, Oklahoma State University, Stillwater OK, United States of America\\
% ...
$^{116}$ LAL, Universit{\'e} Paris-Sud and CNRS/IN2P3, Orsay, France\\
% ...
$^{159}$ Department of Physics, University of Toronto, Toronto ON, Canada\\
% ...
\end{flushleft}


\end{document}